\pdfoutput=1
\documentclass[sigconf]{acmart}

\usepackage{booktabs} 
\usepackage{diagbox}
\usepackage{hyperref}

\setcopyright{rightsretained}

\acmDOI{10.475/123_4}

\acmISBN{123-4567-24-567/08/06}

\acmConference[Submitted to EuroMPI/USA'18]{EuroMPI/USA 2018}{September 23-26, 2018}{Barcelona, Spain}
\acmYear{2018}
\copyrightyear{2018}

\acmArticle{4}
\acmPrice{15.00}

\begin{document}
\title{Performance Evaluation of an Algorithm-based Asynchronous Checkpoint-Restart Fault Tolerant Application Using Mixed MPI/GPI-2}

\author{Adrian Bazaga}
\orcid{0000-0002-1508-285X}
\affiliation{%
  \institution{Barcelona Supercomputing Center (BSC)}
  \streetaddress{C/Jordi Girona, 29-31}
  \city{Barcelona}
  \state{Spain}
  \postcode{E-08034}
}
\email{adrian.bazaga@bsc.es}

\author{Michal Pitonak}
\affiliation{%
  \institution{Computing Center, Slovak Academy of Sciences}
  \streetaddress{Dubravska cesta 9}
  \city{Bratislava}
  \state{Slovak Republic}
  \postcode{845 35}
}
\email{michal.pitonak@savba.sk}

\renewcommand{\shortauthors}{Bazaga et al.}
\renewcommand{\shorttitle}{Bazaga et al.}

\begin{abstract}
One of the hardest challenges of the current Big Data landscape is the lack of ability to process huge volumes of information in an acceptable time. The goal of this work, is to ascertain if it is useful to use typical Big Data tools to solve High Performance Computing problems, by exploring and comparing a distributed computing framework implemented on a commodity cluster architecture: the experiment will depend on the computational time required using tools such as Apache Spark. This will be compared to "equivalent more traditional" approaches such as using a distributed memory model with MPI on a distributed file system such as HDFS (Hadoop Distributed File System) and native C libraries that create an interface to encapsulate this file system functionalities, and using the GPI-2 implementation for the GASPI protocol and it's in-memory checkpointing library to provide an application with Fault Tolerance features. To be more precise, we've chosen the K-means algorithm as experiment, that will be ran on variable size datasets, and then we will compare the computational run time and time resilience of both approaches.
\end{abstract}

%
%



\maketitle

\section{Introduction}

Despite the fact that Apache Spark applications written runs on top of a JVM (Java Virtual Machine), thus can hardly match the FPO performance of Fortran/C(++) MPI \cite{Graham2005} programs compiled to machine code, it has many desirable features of (distributed) parallel application: fault-tolerance, node-aware distributed storage, caching or automated memory management, which are some of the main concerns that arise when dealing with systems resilience \cite{Cappello2009}. Yet we are curious about the limits of the performance of Apache Spark application in High Performance Computing problems by, e.g. writing referential code in C++ and perform comparisons. We do not expect the resulting code to be, in terms of performance, truly competitive with MPI in production applications. Still, such experiment may be valuable for engineers and programmers from the Big Data world implementing, often computationally demanding, algorithms, such as Machine Learning or Clustering algorithms, and also, as with the arrival of the first wave of pre-Exascale machines, the chances for unexpected failures when dealing with highly computationally expensive algorithms during the execution of parallel applications will considerably increase \cite{Bland2013}, it's mandatory to analyze a variety of alternatives to deal with such problems.

The contributions of this work are the following:

\begin{itemize}
\item We design and implement a fault-tolerant application using a mixed MPI/GPI-2 approach, based on checkpoint/restart idea using GPI checkpointing library.
\item We show an example MPI/GPI-2 application implementation that benefits from fault-tolerance features by using the GASPI API, while preserving it's performance, and compare it with a Apache Spark implementation of the same algorithm. 
\item We analyze the additional overhead added by the checkpointing process, depending on the number of nodes that is involved on distributing the mirror data.
\end{itemize}

\section{Background}
\subsection{Apache Spark vs MPI}

In the same way it was mentioned in previous works \cite{Anderson2017}, it has been widely known that High Performance Computing frameworks based on MPI outrun Apache Spark or HDFS-based Big Data frameworks by, usually, more than an order of magnitude for a variety of different application domains, e.g., K-means clustering \cite{Jha2014} and/or large scale matrix factorizations \cite{Gittens2016}. 



It has been shown that it is possible to extend pure MPI-based applications to be elastic in the number of nodes \cite{Raveendran2011} using periodic data redistribution among required MPI ranks. However, this assumes that we are still using MPI as the programming framework, hence we do not get the significant benefits of processing with Apache Spark required for large scale data, such as node fault tolerance. 

Unlike previous approaches, we propose a fair comparison, with similar implementations of algorithms in Spark on HDFS and in C++ on MPI, feeding the algorithms in both cases with distributed data along commodity cluster architectures inside a distributed file system, using native C libraries that create an interface to encapsulate the file systems functionalities, trying to preserve with this the node fault tolerance characteristic exhibited by Spark with HDFS implementations using a one-sided communication approach. We also use HDFS in both approaches to facilitate the storage of large amounts of data without a need for (prohibitive) data replication or sophisticated algorithm-dependent data partitioning. 

\subsection{Fault Tolerance approaches in HPC with MPI}

Several methods have been proposed to deal with Fault Tolerance with MPI. One of the most common approaches to deal with failures in HPC systems are based on rollback-recovery mechanisms \cite{DeCamargo2017}. These strategies allow applications to recover from failures without losing previously computed results. Message logging is a class of rollback-recovery technique that unlike coordinated checkpoint strategies does not require all processes to coordinate to save their state during normal execution and to restart after a single process failure. This means that data is saved regularly so that it can be restored upon failure. For instance, FTI \cite{Bautista-Gomez2011} and SCR \cite{Mohror2014} are popular check-pointing libraries. In \cite{Hursey2011}, a enhancement for a two-phase commit algorithm is explored, by means of changing the linear broadcast and gathering operations with tree-based, log-scaling procedures, they achieve to reduce the complexity. Also, additional error handling mechanisms required to preserve the fault tolerance guarantees of uniform consensus, even with failures, is described.

Moreover, in \cite{Ho2008}, a practical and flexible solution based on group-based coordinated checkpointing and message logging is proposed and implemented. This implementation does not require global checkpoints and restarts, that benefits from a reduced coordination time, and shows a high scalability. Furthermore, in \cite{Gropp2004}, a survey of the features available in the MPI Standard for writing fault-tolerant programs is made, paying focus to different approaches, such as checkpointing, restructuring a class of standard MPI programs, modifying MPI semantics, and extending the MPI specification, which can, within certain constraints, provide a useful context for developing applications that have a significant level of fault tolerance.

Next, in \cite{Elnozahy2002}, the authors review and compare different approaches to rollback-recovery. These approaches can be classified into two broad categories: checkpoint-based protocols and log-based recovery protocols. As the authors demonstrate, coordinated checkpointing generally simplifies recovery and garbage collection, and leads to a good performance in practice, whereas uncoordinated checkpointing does not require the processes to coordinate their checkpoints, thus complicating the recovery, and suffering from problems such as the domino effect. In \cite{Chen2005} it's presented how to build fault tolerant applications by using the FT-MPI with a coding approach. On top of this idea, a floating-point arithmetic version of the Reed-Solomon coding scheme is implemented, and evaluated for performance, yielding a good degree of survivability with a small number of simultaneous node failures, while also preserving low performance overhead and little numerical impact.

Also, in \cite{Chakravorty2006}, a new technique for proactive fault tolerance in MPI applications is presented. This technique is based on the use of task mitigation and load balancing, detecting an imminent fault and attempting to migrate the execution off the processor before a crash happens. In \cite{Bouteiller2015}, an algorithm-based fault tolerance approach for protecting one-sided matrix factorizations is proposed. The authors propose a highly scalable checkpointing method that shows minimal overhead by strategically coalescing checkpoints of many iterations. Also, the accuracy of the method when multiple failures arise is investigated, yielding promising results.

Furthermore, in \cite{Hursey2011a}, a fault tolerant MPI application using a ring communication topology is presented. The authors showcase some of the issues that arise on algorithm-based fault tolerance applications, such as the communication-level issues. Also \cite{Buntinas2008} presents a new implementation of a blocking, coordinated checkpointing, and fault tolerant protocol inside MPICH2, which is evaluated on several high performance architectures, and compare it with Vcl, an implementation of a non-blocking, coordinated checkpointing protocol. The authors demonstrate that, for high speed networks, the blocking implementation gives the best performance for sensible checkpoint frequency, while on other architectures: clusters and computational grids, the checkpointing wave adds significant overhead that doesn't appear in non-blocking approaches.

In \cite{Fagg2004}, the design and use of a fault tolerant MPI that handles process failures in a enhanced way that of the original MPI static process model, is discussed. After that, a short discussion on the consequences of building a fault tolerant MPI, in terms of how such implementations handle failures.In \cite{Ali2014}, a fault-tolerant implementation of a realistic application capable of surviving multiple process failures is described. This implementation takes benefit of the OpenMPI beta process Fault Tolerance specification. From the experimental results, it results clear that it is possible to implement a fault-tolerant application capable of surviving multiple failures by using fault-tolerant OpenMPI, with an acceptable times for gathering the failed processes information and reconstructing the faulty communicator.

\section{Proposed Fault Tolerance approach: Tools \& Libraries}

In this section we describe some of the tools, in a theoretical way, which are used in the present work to provide our non-Spark applications with a Fault Tolerance features.

\subsection{GASPI}

GASPI \cite{Alrutz2013} is a intercommunication framework intended to be used with C++. This framework makes use of the Partitioned Global Address Space (PGAS) communication standard. In this model, each process retains a partition of a globally accessible space in memory. PGAS (Partitioned Global Address Space) programming models have been discussed as an alternative to MPI for some time \cite{Breitbart2014}. The PGAS approach offers the developer an abstract shared address space which simplifies the programming task and at the same time facilitates: data-locality, thread-based programming and asynchronous communication.

\subsection{GPI-2}

GPI-2 \cite{Grunewald2013} is the GASPI implementation, which aims to benefit from the hardware capabilities to use remote direct memory access (RDMA). One of the main focus of GPI-2 is to offer an asynchronous communication in order to overlap the computation and the communication parts, while keeping a thread-safe communication. The API defined by GASPI, comprises a set of one-sided communication procedures, passive-mode communication, global atomics and collectives. On the other hand, the API uses a concept of group, which is intended to imitate the concept of communicator used in MPI, and that is used for the collective operations. Moreover, the API handles the concept of segments, which are blocks of memory contiguous between them, and accessible, for both reading and writing, to the threads working on all the ranks of a GASPI-based application. Thus, the information intended to be exchanged, is stored in the mentioned segments.

The GASPI model was built with the importance of building fault-tolerant applications in mind. The model is able to deal with application-driven fault tolerance on the process level \cite{Shahzad2015}, thus the failure of a particular process does not cause the rest of the application to fail. To do so, the application has the ability to react to a specific failure and restore the computation. To achieve this, GASPI offers two tools: the timeout mechanism for all the blocking procedures, and, since only a timeout mechanism is not enough to discern between whether there has been a failure or not, the concept of error state vector is introduced. The error state vector is a structure that reports the state of each of the processes. A process can be "healthy" (with a \textit{GASPI\_STATE\_HEALTHY} state) or "corrupt" (with a \textit{GASPI\_STATE\_CORRUPT} state). Thus, the applications built using the GASPI model make use of this state vector to determinate the state of remote processes in case of a suspected timeout or error.

Along with other mechanisms available in GPI-2, for the goal of our work, we have used the proposed ping-based Fault Tolerance extension to GPI-2 proposed in \cite{Shahzad2015}, in order to compete with Apache Spark Fault Tolerance features using mixed MPI/GPI2-based applications.

\section{Computational experiment: K-means algorithm}

The K-means clustering \cite{Hartigan1979,Jain2010} is a technique commonly used in machine learning to organize observations into k sets, or clusters, which are representative of the set of observations at large. Observations (S) are represented as n-dimensional vectors, and the output of the algorithm is a set of k n-dimensional cluster centers (not necessarily elements of the original data set) that characterize the observations. Cluster centers are chosen to minimize the within-cluster sum of squares, or the sum of the distance squared to each observation in the cluster:

$$min \sum_{i=1}^{k} \sum_{\vec{x}_{j} \in S_{i}}^{} || \vec{x}_{j} - \vec{\mu}_{i} ||$$

where $S_{i}$ is the set of observations in the cluster i and ${\mu}_{i}$ is the
mean of observations in $S_{i}$.

This problem is NP-hard and can be exactly solved with complexity O(n\textsuperscript{dk+1} log $n$). In practice, approximation algorithms are commonly used to get results that are accurate to within a given threshold by terminating before finally converging, but these algorithms can still take a significant amount of time for large datasets when many clusters are required, and even more when the dimensionality increases.

 The main steps of K- means algorithm are as follows: \newline

\begin{enumerate}
\item Select an initial partition with $K$ clusters; repeat steps 2 and 3 until cluster membership stabilizes.
\item Generate a new partition by assigning each pattern to its closest cluster center.
\item Compute new cluster centers.
\end{enumerate}

\subsection{Parallel approach}

\subsubsection{Method 1: splitting the centers}

Our first parallel K-means implementation aims to parallelize the K-means clustering algorithm by distributing responsibility for finding and assigning the nearest center for each sample to the owner of the center that the sample is currently assigned to. This approach will allow each process to operate on a subset of the samples, effectively dividing the workload. In this method, each processor owns a set of centers that it controls exclusively. After samples are assigned to new centers, assignments to centers that belong to another process are sent over the network to transfer ownership. A collective reduce operation is used to determine whether any assignments have changed, terminating the loop if none have changed. Then, since every process has a set of centers and all associated samples, averaging the samples to calculate the new centers is done entirely independently of every other process. The new centers are then broadcast from every owner process to every other process so that they can be used during the next iteration for determining whether any samples change ownership. The network activity for this method depends on the number of samples that are re-assigned in each iteration. If the ownership for many samples is transferred between nodes, more information needs to be sent over the network. Each of these messages has a single destination, however, so with a good spatial distribution the total bandwidth required should be minimal. The number of ownership changes is also expected to decrease with the number of iterations, since fewer centers should be reassigned as the cluster centers converge to their optimal positions. Sending ownership information requires two unsigned integers for each ownership change. Broadcasting the updated cluster centers requires bandwidth proportional to the total size of the cluster center vectors times the number of processes, since this information needs to be sent between all processes. In this method each processor only has to take care of $\frac{k'}{N}$ centers and preserves in memory n samples, where k' $\leq$ K (equals in the case of only 1 task), N = number of processors and n = number of samples.

\subsubsection{Method 2: splitting the samples}

The second approach parallelizes k-means in a slightly different manner, by assigning ownership of samples to each process instead of distributing the centers. In this case, the assignments are fixed and do not change depending on the center a sample is assigned to. Centers are maintained collectively between processes. During the assignment phase, each sample is assigned to the nearest center. Like the first method, a collective reduce operation is used to determine whether any assignments have changed, terminating the loop if none have changed. A collective reduce operation is used to calculate the new centers by accumulating the samples assigned to each center in each process and using a collective reduce operation to sum the samples for each center across all processes. Another reduce operation is used to globally sum the number of samples assigned to each center. Given this information, each process can independently divide the accumulated samples by the number of samples assigned to each center to obtain the new centers. The network activity for this method is constant and requires transmitting every center vector to every process at the end of each iteration. The amount of bandwidth used is the total size of the cluster center vectors times the number of processes plus the size of the center sample count (an unsigned integer) times the number of cluster centers times the number of processors. In this method each processor only has to take care of $\frac{n'}{N}$ samples and preserves in memory k centers, where n' $\leq$ N (equals in the case of only 1 task), N = number of processors and k = number of centers.

\subsubsection{Apache Spark implementation}

\begin{figure}[tb]
\centering
   \includegraphics[scale=0.35] {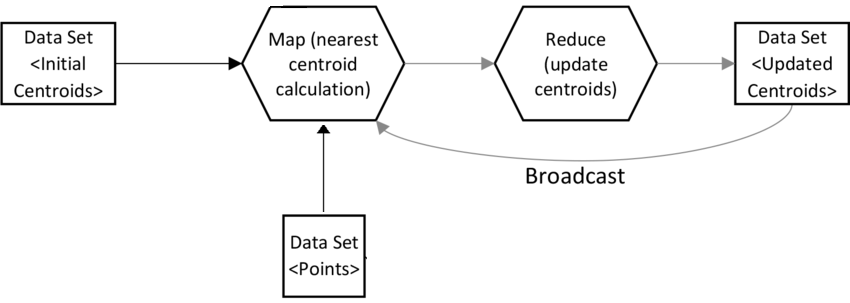}
\caption{Apache Spark K-means data flow graph}\label{kmeansdataflow1}
\end{figure}

In Apache Spark MLib, the K-means algorithm calculates at each iteration a new set of cluster centroids, which is broadcasted back to the first step of the algorithm iterative process. Then the algorithm divides the points' set in several mapping tasks, using the entire centroids' set in each of the tasks. The mapping tasks provides the assignation for the points it owns to their nearest centroids, using a distance algorithm (e.g. Euclidean distance). Finally the mean of the points is computed with a reduce sum procedure on every cluster centroid to obtain a new set of centroids, afterwards the new set is broadcasted to be used in the next iteration. The data flow graph of K-means implementation is depicted in \textit{Figure \ref{kmeansdataflow1}}.

\section{Spark vs MPI vs Mixed MPI/GPI-2 implementations}

In this section, we discuss the different work flows of the different proposed implementations. In every method the data is previously fetched from the Hadoop File System, natively in Apache Spark and through the C++ API in the case of MPI and Mixed MPI/GPI-2 approaches.

\subsection{Apache Spark}

\begin{figure}[tb]
\centering
   \includegraphics[scale=0.32] {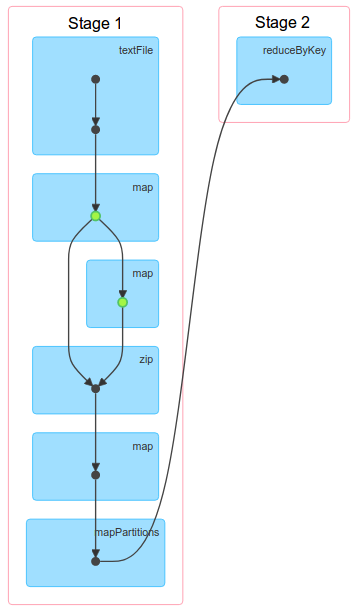}
\caption{Directed Acyclic Graph for the K-means main job in Spark.}\label{sparkdagjobv1}
\end{figure}

In the Apache Spark implementation, each of the K-means iterations is packed in one job, and that job is divided in two stages. In \textit{Figure \ref{sparkdagjobv1}} we illustrate the Directed Acyclic Graph for the K-means main job, with the two stages and their tasks.

\begin{figure*}[tb]
\centering
   \includegraphics[scale=0.3] {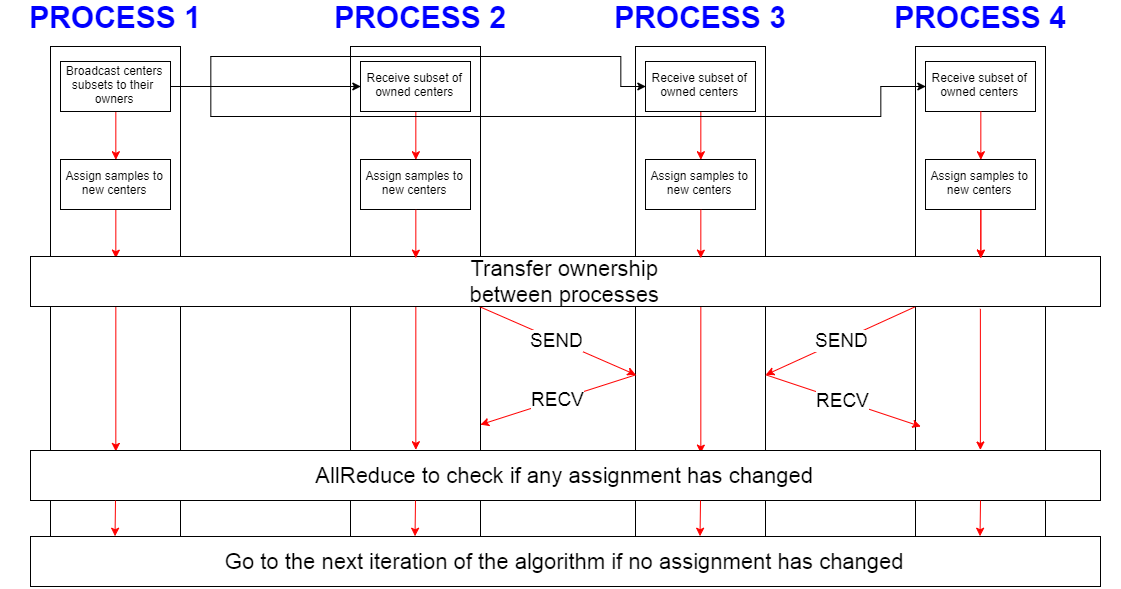}
\caption{Execution diagram of the first MPI method implementation for K-means.}\label{kmeansmpimethod1}
\end{figure*}

The first stage is the map partitions task, where the input data is mapped, this means that with this step we convert each element of the source RDD into a single element of the result RDD by applying a split function to get the correct format to load the algorithm data with it. After that the data is zipped and finally a \textit{mapPartitions} is applied, which converts each partition of the source RDD into multiple elements of the result.

The second stage is a reduction phase, where the output from the first stage is used to generate a new RDD where all values for a single key are combined into a tuple - the key and the result of executing a reduce function against all values associated with that key.

\subsection{MPI}

In the first MPI approach for K-means, we start by calculating the corresponding subset of centers for each process, in order to perform a broadcast after that. After the processes are initialized with their subset of centers and access to the set of samples, they start to assign samples to new centers. If one process assigns a sample to a center which it doesn't owns, the ownership is transferred through the network, in order for each process to only handle their own centers and associated samples.

In \textit{Figure \ref{kmeansmpimethod1}} we depict how process 2 needs to transfer the ownership of some samples to process 3, which owns the centers associated with that samples, the same thing happens with process 4, that needs to transfer the ownership of samples assigned to centers owned by process 3.

After all transfers have concluded, a collective reduce operation is performed in order if any assignment has changed, this way we check if we need to continue iterating in the algorithm or, in the other case, we have finished.

\begin{figure*}[tb]
\centering
   \includegraphics[scale=0.35] {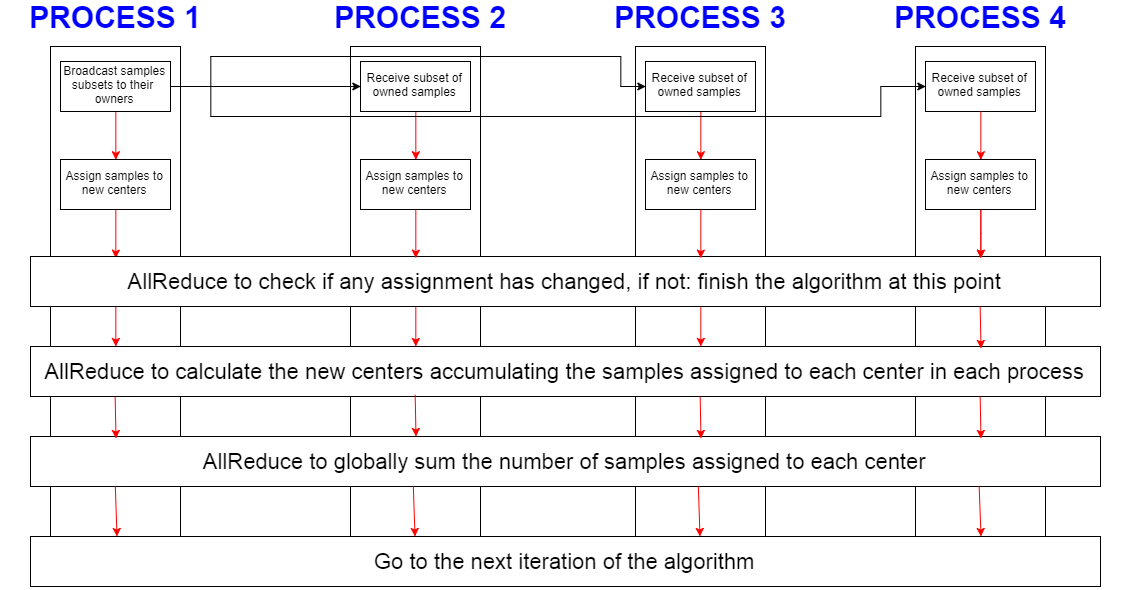}
\caption{Execution diagram of the second MPI method implementation for K-means.}\label{kmeansmpimethod2}
\end{figure*}

In the second MPI approach for K-means, depicted in \textit{Figure \ref{kmeansmpimethod2}}, we start by calculating the corresponding subset of samples for each process, in order to perform a broadcast after that. After the processes are initialized with their subset of samples and access to the set of centers, they start to assign the samples they own to new centers.

After the assignment phase has finished, a collective reduce operation is performed in order if any assignment has changed, this way we check if we need to continue to the next steps or, in the other case, we need to finish the algorithm at the current point.

If any assignment has changed, a collective reduce operation is performed to calculate the new centers for the each subset of samples for each process, to do so we need to accumulate the aforementioned samples.

After the new centers are calculated, another collective reduce operation is carried out to globally sum the number of samples that are assigned to each center, before moving on to the next iteration of the algorithm.

\subsection{Mixed MPI/GPI-2}

In order to have an asynchronous fault-tolerant application, we have used the GPI In-memory checkpoint library \cite{10.1007/978-3-319-64203-1_36}, in order to use a checkpoint/restart based methodology, saving the state of the execution at certain points of it, to be able to recover that state in case of failure. The application needs to decide when it is more reasonable to perform a checkpoint. It must also detect a failure and go into a recovery process. In this recovery process, a spare node takes the position of the node that reported a failure, the last saved checkpoint (state) is read and the application loads it in order to continue it's execution without exiting unsatisfactorily. To have this feature, the application needs to start with a set of available spare nodes, the more spare nodes, the higher tolerance to fails the application will show, since it will have more available nodes on which rely when a failure occurs. This spare nodes are a set of nodes that at a initial stage are idle while the rest of the nodes are executing the algorithm.

In the initialization phase, the application provides a segment, offset and a size where the backup data will be placed in each checkpoint iteration. This is application specific and, in our case, we save the current K-means cluster assignments as backup data. 

\begin{figure}[tb]
\centering
   \includegraphics[scale=0.7] {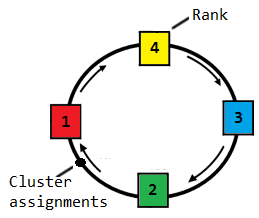}
\caption{Group check-pointing policy overview for a case of 4 ranks. Every rank stores a mirror of it's cluster assignments in the left neighbor. For instance, rank 2 sends it's cluster assignments to be saved by rank 1; rank 1 send it's own cluster assignments to rank 4; rank 4 to rank 3; and finally, rank 3 mirrors it's assignments into rank 2, completing the ring topology.}\label{ranks1v2}
\end{figure}

Also, a checkpoint policy and group must be defined. A group is a GASPI construct formed by a group of nodes (ranks) that will be active and will be able to create checkpoints. The mentioned checkpoint policy is used to define the selection of the neighbor where the mirrored data will be stored. In our case we use a simple ring topology in one direction, where the mirror is always placed on the left or on the right of a node (see \textit{Figure \ref{ranks1v2}} for a general overview of the check-pointing policy).

In the in-memory checkpoint approach we are using, we follow an asynchronous, coordinated checkpoint approach. The coordination is achieved by ensuring global consistency of a snapshot using a GASPI barrier collective operation. The goal of this barrier is, basically, to ensure that there's at least one particular snapshot that is consistent on all processes. The asynchronous part is given by the GASPI communication, so when a checkpoint is performed, the saved data is transferred to the mirror using asynchronous communication.

Doing a checkpoint is a procedure formed of two steps: first you need to start a checkpoint, and second you need to commit it. To commit a checkpoint a global operation is used, ensuring the completion of a previously initiated checkpoint operation on all nodes. At the current point, a valid snapshot now exists, so the application can return to it if required. This commit operation has timeout to avoid blocking. After we've all the GPI-2 environment settled up, we define the two main ad-hoc operations: fault detection and recovery process.

\begin{figure}[tb]
\centering
   \includegraphics[scale=0.38] {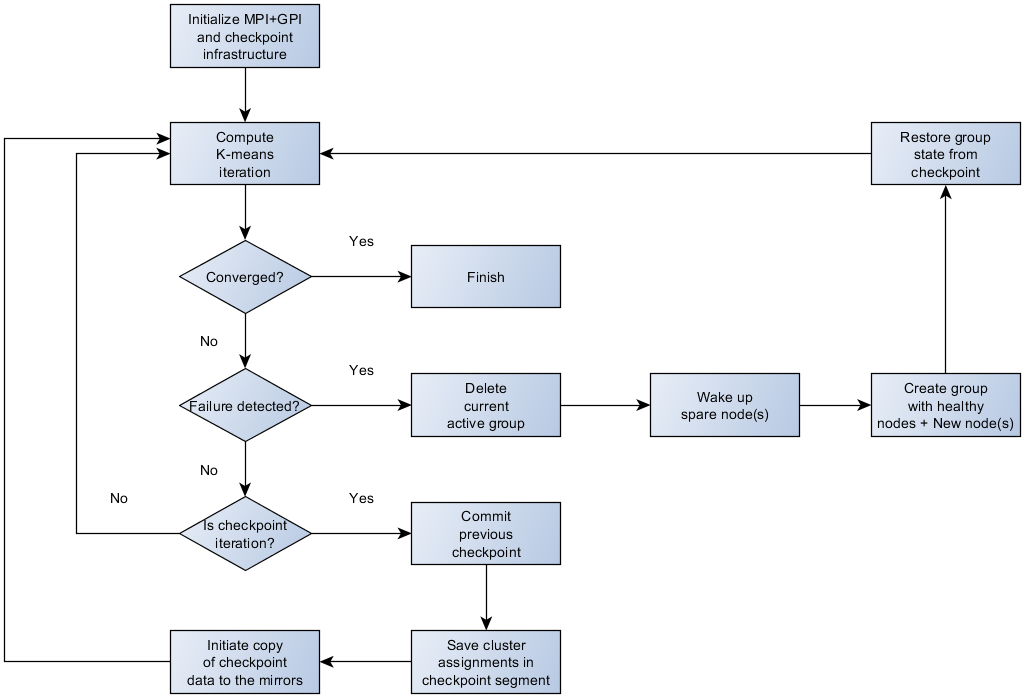}
\caption{Flow diagram for the implemented K-means fault-tolerant implementation.}\label{algorithmkmeansflow}
\end{figure}

The fault detection method we've implemented, relies on a global GASPI barrier operation after each algorithm iteration. This barrier waits for all processes to reach the end of an iteration before reaching a timeout. If some node doesn't make it to this point then we have a failure. In order to detect which nodes have failed and which not, we retrieve the GASPI ranks state vector, which returns the status of each node: healthy or corrupt. After the problem is detected, the failure is communicated to all other running processes, so that all the remaining healthy processes can enter in a consistent way to the recovery process. This idea is depicted on the flow diagram in \textit{Figure \ref{algorithmkmeansflow}}.

The application starts by initializing the MPI and GPI environment, and the GPI checkpoint infrastructure: checkpoint policy, active group of nodes and spare node(s). Then it starts the computation of K-means. If the algorithm has converged, the application finishes, otherwise it checks for a failure on any of the active group nodes by sending a heartbeat check. If any failure is detected, the application deletes the current active group and wakes up the required spare node(s) to substitute the failed ones. Then, a new group comprised of the healthy nodes and the new node(s) is created, where the state is restored from the last committed checkpoint. And the algorithm continues from the first step. If a failure is not detected, the application checks for whether the current iteration is a checkpoint iteration. If it's not, then it continues to the next algorithm iteration. If it's a checkpoint iteration, the application commits the previous checkpoint, saves the cluster assignments of each active node in the GASPI checkpoint segment and initiates the copy of the data to the mirror nodes. Then the application goes to the next iteration of the algorithm.

The recovery process is divided into 3 actions, starting with bringing up a spare node (or more than one if needed) to take the place of the failed one(s). After that we delete the current GASPI group and create a new one, that will be formed by the set of healthy processes and the spare node(s). Now we just need to restore the data, which are the cluster assignments, from the consistent checkpoint saved at the GASPI working segment. In our approach we perform a checkpoint every n iterations of the algorithm, as a simple but reliable benchmarking test, but the checkpoints can we performed using any other criteria.

\begin{figure*}[tb]
\centering
   \includegraphics[scale=0.22] {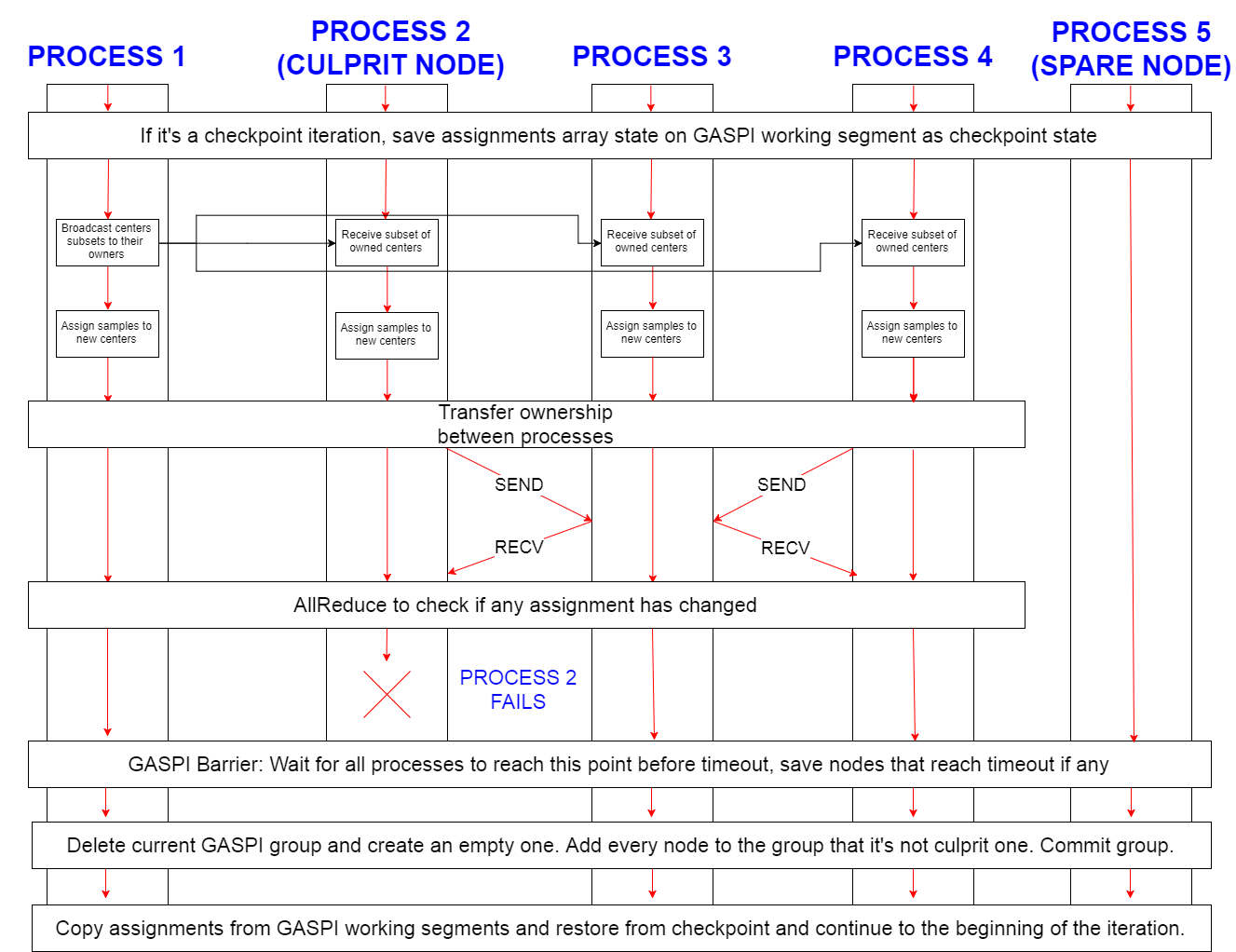}
\caption{Execution diagram of the mixed MPI/GPI-2 implementation for K-means.}\label{kmeansgpi}
\end{figure*}

In \textit{Figure \ref{kmeansgpi}} this mixed MPI/GPI-2 is depicted, with an example iteration where process 2 fails and is overtaken by a spare node, which is process 5.

\section{Performance evaluation}

For benchmarking the execution times of the MPI/GPI-2 application versus Apache Spark, we have used a 1 million 10-dimensional points dataset. We have used 4, 8, 12 and 16 nodes for this benchmarking, but this can be scalated to a much higher amount of nodes. The MPI/GPI-2 application is using 16, 32, 48 and 64 processes for the cases of 4, 8, 12 and 16 nodes, respectively.

The experiment was ran for different values of k, ranging from 10 to 1000 centroids. In \textit{Table \ref{table1}} and \textit{Table \ref{table2}} we have a quantitative performance comparison between Apache Spark and the MPI/GPI-2 application, with the second one overtaking Apache Spark (783 seconds) with a 18x speed-up (37 seconds) in the most computational expensive scenario, which is k = 1000 centroids, using 16 nodes (thus, 64 processes in the case of MPI/GPI-2). Apache Spark yields a execution time significantly slower, with a decrease in speed, on average, of 4 times in comparison with our approach. For the case of Apache Spark, with 4 nodes, the application takes 30, 219 and 783 seconds when asking for 10, 100 and 1000 centroids, respectively. Whereas our approach takes 4, 24 and 150 seconds, respectively. Moreover, for the case of 8 nodes, Apache Spark takes 22, 191 and 720 seconds, while our approach spends 3, 17 and 72 seconds. Next, for the case of 12 nodes, Apache Spark takes 17, 175 and 696 seconds, and our approach yields a execution time of 2, 14 and 50 seconds. Lastly, for the case of 16 nodes, Apache Spark is taking 14, 162 and 650 seconds, whereas our approach finishes in 2, 9 and 37 seconds.

\begin{table}
\centering
\begin{tabular}{ | l | l | l | l |}
    \hline
    \backslashbox{\textbf{Nodes}}{\textbf{k}} & \textbf{10} & \textbf{100} & \textbf{1000} \\ \hline
    \textbf{4} & 30 & 219 & 783 \\ \hline
    \textbf{8} & 22 & 191 & 720 \\ \hline
    \textbf{12} & 17 & 175 & 696 \\ \hline
    \textbf{16} & 14 & 162 & 650 \\
    \hline
    \end{tabular}
    \caption{Execution time (seconds) of the Apache Spark implementation with different number of nodes and a varying number of centroids (k).}\label{table1}
    \end{table}
    
    \begin{table}
\centering
\begin{tabular}{ | l | l | l | l |}
    \hline
    \backslashbox{\textbf{Nodes}}{\textbf{k}} & \textbf{10} & \textbf{100} & \textbf{1000} \\ \hline
    \textbf{4} & 4 & 24 & 150 \\ \hline
    \textbf{8} & 3 & 17 & 72 \\ \hline
    \textbf{12} & 2 & 14 & 50 \\ \hline
    \textbf{16} & 2 & 9 & 37 \\
    \hline
    \end{tabular}
    \caption{Execution time (seconds) of the Mixed MPI/GPI-2 implementation with different number of nodes and a varying number of centroids (k).}\label{table2}
    \end{table}
    
    \begin{figure}[tb]
\centering
   \includegraphics[scale=0.27] {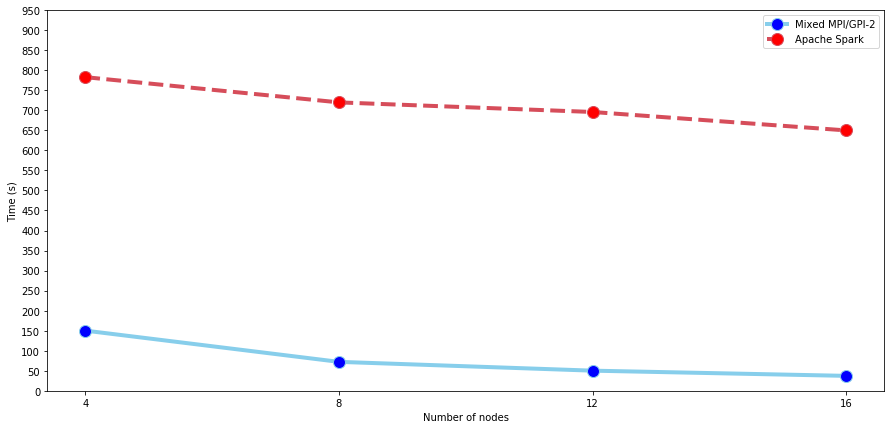}
\caption{Benchmarking of the execution time of K-means on mixed MPI/GPI-2 and Apache Spark on a varying number of nodes with 1 million 10-dimensional points with k = 1000 centroids.}\label{benchmark-executiontime-1}
\end{figure}

\textit{Figure \ref{benchmark-executiontime-1}} shows the execution times of Apache Spark versus our MPI/GPI-2 implementation in the a stressed scenario (k = 1000 centroids). The computation times are the execution times of the applications without taking into account the initialization and the finalization of the programs, measuring exclusively the computation and communication cycles.

\begin{figure}[tb]
\centering
   \includegraphics[scale=0.27] {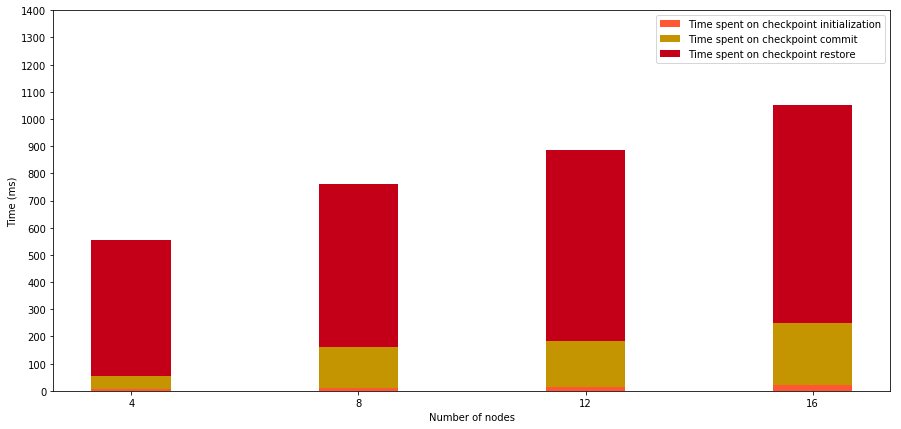}
\caption{Overhead of the checkpointing process when performing the three main actions, every 50 iterations of the algorithm, measured for the case of k = 1000. Waiting time is depicted in milliseconds.}\label{benchmark-cp-1}
\end{figure}

\textit{Figure \ref{benchmark-cp-1}} shows the time spend by the different phases of the in-memory checkpointing library versus the number of nodes. The waiting times for the checkpointing procedures range from 550 milliseconds, in the case of 4 nodes, to 1100 milliseconds, in the case of 16 nodes. The checkpoints are made in a cycle every 50 iterations, thus, since the convergence for the case of k = 1000 takes 550 iterations, 11 checkpoints have been written during the execution of the application. The time spent in the checkpoint initialization is relatively small. The commit procedure, that asserts if the checkpoint has already been completed, is not dependent of the number of nodes, if the gap between one checkpoint and the next one is chosen in such a way that the underlying asynchronous operations are able to execute.

Consequently, the in-memory GPI-2 library features don't add significant delay in the execution due to the fact it uses, at the logical level, the GASPI asynchronous methodology to perform all the checkpoint savings and fault detection. Moreover, we can also conclude that the time spent on checkpoints on a failure scenario, is less than 1\% of the total computation time taken by the algorithm to converge, thus is not adding a major overhead.

\section{Conclusion}
This study has presented the implementation of an asynchronous fault-tolerant K-means algorithm, mixing MPI and GPI-2 features. We have implemented two different parallelization approaches for the suggested algorithm, in the first one, we parallelize the computation by splitting the center assignments, and in the second one, the samples are distributed among the ranks. Moreover, on top of the MPI implementation, we added a fault-tolerance layer based on the GASPI protocol and the GPI-2 in-memory checkpointing library, that takes care of the checkpointing policy in a periodic basis, and restores the computation in case of failure.

Moreover, we have demonstrated that MPI/GPI-2 approaches are able to compete with Apache Spark to solve High Performance Computing problems, while preserving Fault Tolerance
features, such as the ones exposed by Apache Spark, by using the in-memory checkpoint capabilities available in the GPI-2 framework.

As future work, we are currently working on studying the performance impact of distributing the workload of the execution between the set of active nodes, without replacing the faulty nodes, since this can become a significantly relevant output for computing centers with limited resources.

\begin{acks}
  This work has been partially funded by the Summer of High Performance Computing program by PRACE (Partnership for Advanced Computing in Europe).

The computing was performed in the High Performance Computing Center of the Slovak Academy of Sciences using the HPC infrastructure acquired in project ITMS 26230120002 and 26210120002 (Slovak Infrastructure for High-Performance Computing) supported by the Research \& Development Operational Programme funded by the ERDF.
\end{acks}


\end{document}